\documentclass[amsmath,amssymb,showpacs,showkeywords]{revtex4}
\usepackage{amsmath,amsfonts,amssymb,latexsym,graphicx,youngtab,extarrows}
\newtheorem{thm}{Theorem}
\newtheorem{lemma}{Lemma}

\newtheorem{rema}{Remark}

\newcommand{\dis}{\displaystyle}

\newcommand{\bequ}{\begin{equation}}
\newcommand{\eequ}{\end{equation}}
\newcommand{\barr}{\begin{array}}
\newcommand{\earr}{\end{array}}
\newcommand{\bea}{\begin {eqnarray}}
\newcommand{\eea}{\end {eqnarray}}
\newcommand{\lb}{\label}

\newcommand{\qed}{\hfill \rule{2.25mm}{2.25mm}\vspace{.15cm}}

\renewcommand{\Re}{{\cal R}{\rm e}\:}
\begin{document}
% ====================================
\let\la=\lambda
\def \Z {\mathbb{Z}}
\def \Zt {\mathbb{Z}_o^4}
\def \R {\mathbb{R}}
\def \C {\mathbb{C}}
\def \La {\Lambda}
\def \ka {\kappa}
\def \vphi {\varphi}
\def \Zd {\Z ^d}
% ====================================
% **************************************************
% **************************************************
%\begin{frontmatter}
\title{On Thermodynamic and Ultraviolet Stability of Yang-Mills}
\author{Paulo A. Faria da Veiga}\email{veiga@icmc.usp.br. Orcid_Registration _Number: 0000-0003-0739-069X.}
\author{Michael O'Carroll}\email{michaelocarroll@gmail.com}
\affiliation{Departamento de Matem\'atica Aplicada e Estat\'{\i}stica - ICMC, USP-S\~ao Carlos,\\C.P. 668, 13560-970 S\~ao Carlos SP, Brazil}
% **************************************************
% **************************************************
% **************************************************
% **************************************************
% **************************************************
% **************************************************
% *******DATA*******DATA******DATA*****DATA*******DATA*****  DATA
\date{October 07, 2019.} %{\bf\large DRAFT VERSION}\vspace{.3cm}}
% **************************************************
% **************************************************
% **************************************************
% **************************************************
% **************************************************
% **************************************************
% **************************************************
% **************************************************
\begin{abstract}
 We prove thermodynamic and ultraviolet stable stability bounds for the pure Yang-Mills relativistic quantum theory in an imaginary-time, functional integral formulation. We consider the gauge groups $\mathcal G={\rm U}(N)$, ${\rm SU}(N)$ and let $d(N)$ denote their Lie algebra dimensions. We start with a finite hypercubic lattice $\Lambda\subset a\mathbb Z^d$, $d=2,3,4$, $a\in(0,1]$, $L\in\mathbb N$ sites on a side, and with free boundary conditions. The Wilson partition function $Z_{\Lambda,a}\equiv  Z_{\Lambda,a,g^2,d}$ is used, where the action is a sum over gauge-invariant plaquette actions with a pre-factor $(a^{d-4}/g^2)$, where $g^2\in(0,g_0^2]$, $0<g_0<\infty$, defines the gauge coupling. Each plaquette action is pointwise positive. Formally, in the continuum limit $a\searrow 0$, this action gives the well-known Yang-Mills action. Either by using the positivity property and neglecting some of the plaquette actions or by fixing an enhanced temporal gauge, which involves gauging away the bond variables belonging to a maximal tree in $\Lambda$, and which does not alter the value of $Z_{\Lambda,a}$, we retain only $\Lambda_r$ bond variables. $\Lambda_r$ is of order $[(d-1)L^d]$, for large $L$. We prove that the normalized partition function $Z^n_{\Lambda,a}=(a^{(d-4)}/g^2)^{d(N)\Lambda_r/2}Z_{\Lambda,a}$ satisfies the stability bounds
 $e^{c_\ell d(N)\Lambda_r}\leq Z^n_{\Lambda,a}\leq e^{c_ud(N)\Lambda_r}$, with finite $c_\ell,\,c_u\in\mathbb R$ independent of $L$, the lattice spacing $a$ and $g^2$. In other words, we have extracted the {\em exact} singular behavior of the finite lattice free-energy. For the normalized free energy  $f_{\Lambda,a}^n=[d(N)\,\Lambda_r]^{-1}\,\ln Z^n_{\Lambda,a}$, our stability bounds imply, at least in the sense of subsequences, that a finite thermodynamic limit  $f^n_a\equiv\lim_{\Lambda\nearrow a\mathbb Z^d} f_{\Lambda,a}^n$ exists. Subsequently, a subsequential finite continuum limit $f^n\equiv \lim_{a\searrow 0}f^n_a$ also exists.
\end{abstract}
\maketitle

%%%%%%%%%%%%%%%%%%%%%%%%%%%%%%&&&&&&&&&&&&&&&&&&&&&&&%%%%%%%%%%%%%%%%%%%%%%%%%%%%%
\section{Introduction and Results} \lb{intro}
%%%%%%%%%%%%%%%%%%%%%%%%%%%%%%&&&&&&&&&&&&&&&&&&&&&&&%%%%%%%%%%%%%%%%%%%%%%%%%%%%%
To show the existence, the particle spectrum, the particle interaction and scattering in a quantum field theory (QFT) are among the most fundamental problems in physics \cite{Wei,Banks}. Unfortunately, in spite of much work and  progress (see e.g. \cite{GJ,Riv,Summers}), we lack a physically relevant, mathematically well-defined QFT in spacetime dimension $d=4$. For many reasons, Quantum Chromodynamics (QCD) seems to be the best candidate for such a model.

In this context, neglecting the Fermi matter fields, the existence of nonabelian pure-gauge, Yang-Mills models was considered in a series of papers \cite{Bal,Bal2} (and Refs. therein) using a lattice regularization for the continuum spacetime \cite{Sei,Gat} and employing intricate analytical tools as e.g. multiscale methods based on the renormalization group, and small/large field decompositions. Within this framework, thermodynamic and ultraviolet stability bounds \cite{Rue} were proven for $d=3,4$. In \cite{BFS}, abelian gauge Bose matter models were considered in $d=2$. More recently, in Ref. \cite{Dim}, abelian gauge models with fermions were considered in $d=3$. But, up to now, stability bounds have {\em not} been proved for gauge-matter models like QCD. Together with confinement, this is a very challenging problem.

Here, we provide a very simple proof of thermodynamic and ultraviolet stability of Yang-Mills in Euclidean dimension $d=2,3,4$ and for abelian/nonabelian connected and compact gauge Lie groups $\mathcal G$. We work in the lattice configuration space and our method is direct and does not employ sophisticated analysis. Instead, it exploits the pointwise positivity of the gauge-invariant Wilson lattice plaquette action \cite{Wil,Gat}, gauge invariance, properties of the Haar measure on $\mathcal G$ \cite{Simon2}, a relation with random matrices and, finally, the Weyl formula \cite{Weyl,Bump} for the integration over the gauge group $\mathcal G$ of class functions (functions with constant values in the conjugacy classes of $\mathcal G$). Our treatment is rigorous, uses the gluon fields and differs from the character representation approach of \cite{Ash}, for $d=2$.

Our analysis applies to other connected compact Lie groups, but we focus on $\mathcal G={\rm U}(N)$, ${\rm SU}(N)$, and denote by $d(N)$ the dimension of their Lie algebras [$N^2$ and $(N^2-1)$, respectively]. For simplicity, we will treat explicitly the case $\mathcal G={\rm U}(N)$. In the end of the paper, we show how the proofs are modified for $\mathcal G={\mathrm SU}(N)$. Our starting point is the finite-lattice partition function. We let $\Lambda\subset a\mathbb Z^d$, $a\in(0,1]$, be a finite hypercubic lattice with $L\in\mathbb N$ sites on each side. A lattice site is denoted by $x=(x^0,x^1,\ldots,x^{d-1})$;  $x^0$ is the time coordinate. If $\nu=0,1,\ldots,(d-1)$ is a coordinate direction and $e^\nu$ its unit vector, $b_\nu(x)$ denotes a lattice nearest-neighbor bond starting at $x$ and ending at $x+ae^\nu\equiv x^\nu_+$, with $x,x^\nu_+\in\Lambda$. To each lattice bond $b_\nu(x)$, we assign a gauge bond variable which is a unitary matrix $g_{b_\nu(x)}\in\mathcal G$. The partition function for our model is given by (see Refs. \cite{Sei,Gat})
\bequ\lb{Z}
Z_{\Lambda,a}\,\equiv\,Z_{\Lambda,a,g,d}\,=\,\dis\int\exp\left[-\frac {a^{d-4}}{g^2}\,\sum_p\,\!\mathcal A_p\right]\;d\tilde g\,.
\eequ
Here, $g^2>0$ defines the pure-gauge coupling, the measure $d\tilde g$ is a product of normalized Haar measures  $d\sigma(g_b)$ on $\mathcal G$, one for each lattice bond $b$, with no distinction of orientation. If $p\in\Lambda$ is a {\em plaquette} in the $\mu-\nu$ plane (a minimal square), $\mu<\nu$, with vertices at the sites $x$, $x_{+}^\mu$, $x_{+}^\mu+ae^\nu$ and $x_{+}^\nu$, then the single plaquette $p$ action is given by
\bequ\lb{action}
\mathcal A_p\,=\,\|1-U_p\|^2_{H-S}\,=\,2\,\Re {\mathrm Tr}(1-U_p)\,,
\eequ
where $U_p\,=\,U_1U_2U_3U_4$, with $U_1=g_{x,x_{+}^\mu}$, $U_2=g_{x_{+}^\mu,x_{+}^\mu+ae^\nu}$, $U_3=g^{-1}_{x_{+}^\nu,x_{+}^\nu+ae^\mu}$ and $U_4=g^{-1}_{x,x_{+}^\nu}$. Here, $g^{-1}$ is the inverse element of $g$ and, for a square matrix $M$ with trace $Tr\,M$ and adjoint $M^\dagger$, $\|M\|_{H-S}=[Tr(M^\dagger M)]^{1/2}$ is the Hilbert-Schmidt norm. We adopt free boundary conditions, which means that we include all the bonds connecting nearest neighbor sites of $\Lambda$. Note that each plaquette action $\mathcal A_p$ is {\em pointwise positive}. Also, we remark that our model verifies Osterwalder-Schrader positivity, for $L$ even \cite{Sei}.

Formally, using the well-known Baker-Campbell-Haussdorff formula and the parametrization $g_b=e^{iagA_b\theta}$, where $A_b$ is the physical gluon field or gauge potential, in Ref. \cite{Gat} it is shown that the Wilson action $(a^{d-4}/g^2)\sum_p \mathcal A_p$ is, for small $a$, the Riemann sum approximation to the usual smooth field classical continuum Yang-Mills action ${\mathrm Tr} F^2$, where $F^{\mu\nu}$ is field strength antisymmetric tensor. 

We now discuss the parametrization of the bond variables for the gauge group $\mathcal G={\rm U}(N)$. (Much of our discussion also applies to ${\rm SU}(N)$.) Fixing a lattice bond $b$ and, for the associated gauge variable $U_b$, we write $U_b=e^{iX_b}$, where $X_b=x^b_{\alpha}\theta_\alpha$, with a sum over $\alpha=1,2,\ldots,d(N)=N^2$. The self-adjoint $N\times N$ matrices $\theta_\alpha$ form a Lie algebra basis and are taken to obey the normalization $Tr\,(\theta_\alpha\theta_\alpha)=\delta_{\alpha\beta}$, with a Kronecker delta. We call the real parameters $x^b_{\alpha}$ the {\em gluon} fields. In terms of this parametrization, the Haar measure for each bond $b$ gauge variable is a product of a density and a $d(N)$-dimensional Lebesgue measure. Whenever $N>2$, we know some global gluon parametrization, as e.g. the parametrizations for ${\rm SU}(N)$ in terms of Euler angles (see e.g Ref. \cite{Euler}). However, in general, it is not clear that we can characterize their domains such that the global parametrization is also an injection between a group element and a parameter value, leading to a good characterization of the Haar measure on the gauge group ${\mathcal G}$. With these limitations, it is difficult to bound integrals of functions on $\mathcal G$. More precisely, in our case e.g. the Euler angle parametrizations do not lend to use good quadratic approximations. (See the end of this section where this point is made more clear.) In the special case of ${\rm SU}(2)$, a global gluon parametrization, its domain and an explicit formula for the density of the measure is available (see Ref. \cite{Simon2}).

In $Z_{Y,\Lambda,a}$, instead of using the {\em physical gluon} potential, if we parametrize the bond variable $g_b\in\mathcal G$ by the fields $X_b$, so that $g_b=e^{iX_b}$, $X_b=\sum_{\alpha=1,\ldots,d(N)} x^b_\alpha \theta_\alpha$, then the corresponding Haar measure is approximately a constant $c$ times the Lebesgue measure $d^{d(N)}x^b$ and
$$
Z_{Y,\Lambda,a}\,\simeq\,\dis\int\,\exp\left\{ -\dfrac{a^{d-4}}{g^2}\sum_p \mathcal A_p(g_b=e^{iX_b};b\in p)\right\}\,\prod_{b}c\,d^{d(N)}x^b\,.
$$

Now, the $X_b$ fields are related to the physical gluon fields $A_b$ by the local scaling relation
$agA_b\,=\,X_b$. We denote by $y_b$ the scaled fields which are related to $A_b$ by the local scaling relation
$$y_b\,=\,a^{(d-2)/2}\,A_b\,.$$
By the above, the scaled gauge field $y_b$ is related to the $X_b$ field by
$$
y_b\,=\,a^{(d-2)/2}\,A_b\,=\,a^{(d-2)/2}\dfrac1{ag}X_b\,=\,\dfrac{a^{(d-4)/2}}{g}\,X_b\,.
$$
Finally, in terms of the $y_b$ fields, the partition function $Z_{Y,\Lambda,a}$ becomes
$$
Z_{Y,\Lambda,a}\,\simeq\,\left[ga^{(4-d)/2}\right]^{d(N)\Lambda_r}\,
\dis\int\,\exp\left\{-\dfrac{a^{d-4}}{g^2}\sum_p \mathcal A_p(g_b=e^{i\frac{g}{a^{(d-4)/2}}y_b\theta})\right\}\,\prod_{b}c\,d^{d(N)}y_b\,.
$$
The exponent of the exponential in the integrand is nonsingular for small $a$ and, for $g^2\leq g_0^2$, $0<g_0<\infty$.

Based on this approximate small field relation, we define a scaled, normalized partition function $Z^n_{Y,\Lambda,a}$ by
\bequ\lb{Znorm}
Z^n_{Y,\Lambda,a}\,=\,\left[\dfrac{a^{d-4}}{g^2}\right]^{d(N)\Lambda_r/2}\,Z_{Y,\Lambda,a}\,.
\eequ
It is precisely this normalized partition function that obeys thermodynamic and ultraviolet stable stability bounds.

We now consider another parametrization which is global for class functions on $\mathcal G$. An arbitrary element $U\in{\rm U}(N)$ (not necessarily near the identity) is unitarily equivalent to a diagonal matrix $D={\rm diag}(e^{i\lambda_1}, \ldots, e^{i\lambda_N})$, $\lambda_j\in(-\pi,\pi]$. We refer to the $\lambda_j$ as the {\em angular eigenvalues}. If we are integrating a class function $f(U)=f(VUV^{-1})$ (for any $U$ and all $V\in{\mathcal G=\rm U}(N)$), then $f(U)$ is a function of the angular eigenvalues only. By the Weyl integration formula \cite{Weyl,Simon2,Bump}, the gauge group integral of $f(U)$ is equal to an integral over the $N$ angular eigenvalues with a measure which is the product of a $N$-dimensional Lebesgue measure and an explicit density function; the integration domain is $(-\pi,\pi]^N$.

The functions we encounter in $Z$ of Eq. (\ref{Z}) are {\em not} class functions of each bond variable. However, the bounds we obtain are class functions so that the Weyl integration formula and the angular eigenvalue parametrization can and is used. This property and the fact the bounds on $Z$ factorize into products of single plaquette partition functions of a single bond variable play an important role in our method.

Let us now discuss local gauge invariance, gauge fixing and the model degrees of freedom. Recall the property of local gauge invariance of $Z_{\Lambda,a}$. The group $\otimes_x \mathcal G$, with element $\prod_x r_x$, acts on bond variables mapping $g_{b_{\mu(x)}}$ to $(r_x g_{b_{\mu(x)}} r^{-1}_{x+ae^\mu})$; each plaquette action $\mathcal A_p$ and the total action $\mathcal A$ are invariant under this mapping. Due to local gauge invariance, there is an excess of bond variables in Eq. (\ref{Z}). By a gauge fixing procedure (see Chap. 22 of Ref. \cite{GJ}), some bond variables can be eliminated  or {\em gauged away} by setting them to the identity in the action, i.e. $g_b=1$, for a bond $b$, and omit the bond integration variable (its integral gives $1$!). The value of the partition function $Z_{\Lambda,a}$ is unchanged in this process, provided the associated bonds do {\em not} form closed loops. As each bond variable has $d(N)$ gluon fields, we reduce the number of degrees of freedom by $d(N)\Lambda_g$, where $\Lambda_g$ is the number of gauged away bonds. We denote by $\Lambda_r$ the number of {\em retained bond} variables and, after the gauge fixing process, we are left with $d(N)\Lambda_r$ gauge field degrees of freedom.

Here, sometimes it is convenient to choose the {\em enhanced temporal (axial)} gauge which we now define. If we identify the sites of the $\mu$-th coordinate with $1,2,\ldots,L$, this gauge is defined by setting the following bond variables to $1$. First, for any $d=2,3,4$, we take $g_{b_0(x)}=1$. For $d=2$, take also $g_{b_1(x^0=1,x^1)}=1$. For $d=2$, the gauged away bonds form a comb with the teeth along the $x^0$-direction. The roots of the teeth lie on $x^0=1$; the open end is $x^0=L$. For $d=3$, set also $g_{b_1(x^0=1,x^1,x^2)}=1$ and $g_{b_2(x^0=1,x^1=1,x^2)}=1$. Similarly, for $d=4$, take also $g_{b_1(x^0=1,x^1,x^2,x^3)}=1$, $g_{b_2(x^0=1,x^1=1,x^2,x^3)}=1$ and $g_{b_3(x^0=1,x^1=1,x^2=1,x^3)}=1$. For $d=3$, the gauged away bonds can be visualized as forming a scrub brush with bristles along the $x^0$ direction and the grip forming a comb. In $d=2,3,4$, the gauged away bonds do not form loops, and there are $\Lambda_r\equiv\Lambda_r(d)$ remaining variables. $\Lambda_r$ has the values $(L-1)^2$, $[(2L+1)(L-1)^2]$, $[(3L^3-L^2-L-1)(L-1)]$, for $d=2,3,4$, respectively. Clearly, $\Lambda_r\simeq (d-1)L\nearrow\infty$, as
$\Lambda\nearrow a\mathbb Z^d$. Note that, fixing the enhanced temporal gauge, the unretained or gauged away gauge variables are associated with bonds in the hypercubic lattice $\Lambda$ which form a {\em maximal tree}, so that, by adding any other bond to it, we form a closed loop. Thus, the partition function is not changed by the process of gauge fixing (see \cite{GJ}).

The main result of this paper is the following thermodynamic and ultraviolet stable, stability bounds.\vspace{-1.0mm}
\begin{thm}
Let $d=2,3,4$, $a\in(0,1]$ and $g^2\in(0,g_0^2]$, with $0<g_0<\infty$. Then, the normalized partition function for our model, with free boundary conditions, \bequ\lb{nZ}Z^n_{\Lambda,a}=(a^{d-4}/g^2)^{d(N)\Lambda_r/2}\;Z_{\Lambda,a}\,,\eequ satisfies the stability bounds
$$\exp[c_\ell d(N)\Lambda_r]\leq Z^n_{\Lambda,a}\leq \exp[c_ud(N)\Lambda_r]\,,$$ with finite constants $c_\ell,\,c_u\in\mathbb R$ independent of $\Lambda_r$ (and, hence, of $\Lambda$), $a$ and $g^2$. In these stability bounds, the exact singular behavior of the finite lattice free energy has been extracted and isolated. At least in the sense of subsequences, a thermodynamic limit  $\Lambda\nearrow a\mathbb Z^d$ of the normalized free energy $$f^n_{\Lambda,a}=[d(N)\,\Lambda_r]^{-1}\,\ln Z^n_{\Lambda,a}$$ exists and is finite. Subsequently, a finite continuum limit $a\searrow 0$ also exists.\vspace{-1.5mm}
\end{thm}
\begin{rema}\lb{rema1}
The existence of the above finite subsequential limits for the free energy follow, first, by considering a sequence of hypercubic lattices, with a fixed lattice spacing $a$ and a fixed lattice $\Lambda$. Taking the sequence of the corresponding normalized free-energies, by the Bolzano-Weierstrass theorem, there is a convergent subsequence in the thermodynamic limit $\Lambda\nearrow a\mathbb Z^d$. Subsequently, we take a sequence of lattices in $a\mathbb Z^d$ and look at their normalized free-energies. Applying the Bolzano-Weierstrass theorem to this sequence shows there is a convergent subsequence in the continuum limit $a\searrow 0$. Here, we do not consider the unicity of the limits, but only their existence and finiteness. Model correlations and physically relevant properties like the correlation decay rates and the model energy-momentum spectrum must be considered separately and will be the subject of a future analysis. Besides the result of Theorem 1, the importance of our method is that it also allows to extend the analysis and obtain stability bounds for gauge models with matter fields \cite{bQCD}. \vspace{-1.5mm}
\end{rema}

The rest of the paper is devoted to the proof of the stability bounds of Theorem 1. With this and using Remark \ref{rema1}, Theorem 1 is proved. 

For the upper stability bound on $Z^n_{\Lambda,a}$, we do not fix a gauge. A special role is played by a {\em single-plaquette partition function with a single bond variable}. For $\mathcal G={\rm U}(N)$ and with $\mathcal N(N)\equiv 1/[(2\pi)^N N!]$, it is given by (the integral is $1$ if all $\lambda_k\equiv 0$)
\bequ\lb{zz}\barr{lllll}
\!\!z(c)&=&\displaystyle\int\,e^{-c\|1-g\|^2_{H-S}}\,d\sigma(g)&=&\mathcal N(N)\,\displaystyle\int_{[-\pi,\pi]^N}\,e^{-c\sum_{k=1}^N 2(1-\cos\lambda_k)}\,\rho(\lambda)d^N\lambda\,.\earr
\eequ
Here, $\lambda=(\lambda_1,\ldots,\lambda_N)$, the integration Haar measure was defined according to Eq. (\ref{Z}), $d^N\lambda=d\lambda_1\ldots d\lambda_N$ and the
density $\rho(\lambda)$ is, for $1\leq j<k\leq N$, the square of a Vandermonde determinant
$$\rho(\lambda)=\prod_{j<k}\left|e^{i\lambda_j}-e^{i\lambda_k}\right|^2=\prod_{j<k}2[1-\cos(\lambda_j-\lambda_k)]\,.$$
The second equality in Eq. (\ref{zz}) comes from the application of the Weyl integration formula for a class function \cite{Weyl,Simon2,Bump}. As defined above, for a $N\times N$ unitary matrix with eigenvalues $e^{i\lambda_1}$, $e^{i\lambda_2}$, ..., $e^{i\lambda_N}$, $\lambda_k\in (-\pi,\pi]$, $\lambda_1$, ...,$\lambda_N$ are the angular eigenvalues. Finally, in Eq. (\ref{zz}), $c\equiv c(a,g^2,d)\in\mathbb R$ is a constant depending on $[a^{d-4}/g^2]$ and will be specified later. An important observation is that the plaquette action $\mathcal A_p=\|1-U_p\|^2_{H-S}$ is {\em not} a class function of each bond variable. Also, for $d=2$, it is known (see \cite{Ash}) that the exact result is $Z_{\Lambda,a}= z^{\Lambda_r}$, and $c=(ga)^{-2}$ and $\Lambda_r$ is equal to the total number of plaquettes in $\Lambda$.

For the lower bound on $Z^n_{\Lambda,a}$, the enhanced temporal gauge is fixed. There is also a characteristic function in the integrand of $z(c)$ of Eq. (\ref{zz}). This function restricts each bond variable $g$ to be close to the identity. Note that the integrand in $Z^n_{\Lambda,a}$ (as in $Z_{\Lambda,a}$) is positive so that restricting the integration domain of the gauge variables gives a lower bound on the integral.

The importance of Eq. (\ref{zz}) is that, by discarding some horizontal plaquettes, $Z_{\Lambda,a}$ is bounded by a product of $z(c)$ (modified with a characteristic function, for the lower bound), and we can bound $Z_{\Lambda,a}$ by bounding $z(c)$. In this process, we remark that although the four-bond plaquette action $\mathcal A_p$ is {\em not} a class function of each bond variable on $\mathcal G$, the integrand of the single bond partition $z(c)$ is, and the Weyl integration formula applies. In turn, $z(c)$ is bounded by bounding the angular eigenvalue distribution. In the bounds, a smaller number of fields appear; namely $N$ angular eigenvalue fields rather than $N^2$ gluon fields.

Before proving the stability bounds, we give some intuition about how our methods work and also the relation between our gluon fields and the usual gauge potentials $A_\mu(x)$, where a gauge group element $g_{b_\mu(x)}$, associated with a lattice bond $b_\mu(x)$, is parametrized as $\exp[iga A_\mu(x)]$. We refer to the gauge potential $A_\mu$ as the {\em physical} gluon field. If the factor  $(ga)$ is not present, which refer to the field as the gluon potential or field. 

In the enhanced temporal gauge, we only have $\Lambda_r$ gauge bond variables left. The group bond variable, in terms of gluon fields, is $e^{ix_\alpha\theta_\alpha}$. Formally, in the quadratic approximation for the action the partition function is given by
$$
Z_{\Lambda,a}\,\simeq\,\int e^{-\frac{a^{d-4}}{g^2}\,\sum_p x_p^2}\,d\tilde x\,,	
$$
where, for  $x_1,\ldots,x_4$ denoting the gluon fields or the gauge group parameters associated with the four consecutive sides of a plaquette $p$, we have set $x_p=x_1 +x_2-x_3-x_4$.

In terms of the {\em scaled fields} $y$ defined by $y_\ell= \dfrac{a^{(d-4)/2}}{g} \,x_\ell$, we have
$$Z_{\Lambda,a}\,\simeq\,\left(\frac{a^{d-4}}{g^2}\right)^{-d(N)\Lambda_r/2}\,\int e^{-\,\sum_p y_p^2}\,d\tilde y\,.$$
Here, note the presence of the $a\searrow 0$ singular scaling factor of $\ln Z_{\Lambda,a}$ appearing in the free energy, which is eliminated in the definition of the normalized partition function on Eq. (\ref{nZ}). Now, in terms of the gauge potentials $A_\mu(x)=A_{\mu,\alpha}(x)\theta_\alpha$, we write the bond variables as $g_{b_\mu(x)}=\exp (igaA_\mu(x))$ and
$$
Z_{\Lambda,a}\,\simeq\, (ag)^{d(N)\Lambda_r}\,\int  e^{-a^{d-2}\,\sum_p A_p^2}\, d\tilde A\,.	
$$
For a plaquette in the $\mu-\nu$ plane, $\mu<\nu$, with a corner at site $x$,
$A_p^2=(A_1+A_2-A_3-A_4)^2=a^2(\partial_\mu A_\nu(x)-\partial_\nu A_\mu(x))^2$, where $\partial_\mu A_\nu(x)$ is the finite difference derivative $\partial_\beta f(x)=[f(x+ae^\beta)-f(x)]/a$.
Hence, we obtain
$$
Z_{\Lambda,a}\,\simeq\, (ag)^{d(N)\Lambda_r}\,\int  e^{-a^{d}\,\sum_{\mu,\nu;\mu<\nu} [\partial_\mu A_\nu(x)-\partial_\nu A_\mu(x)]^2}\, d\tilde A\,,
$$
so that the exponent is the Riemann sum approximation to the quadratic part of the classical (nonabelian) Yang-Mills action. The relation between the components of $y$ fields and the $A$ fields is $y=a^{(d-2)/2}A$. Based on these arguments, for $Z_{\Lambda,a}$ written in terms of the physical gluon fields $A_\mu$, we define the normalized partition function $Z^n_{\Lambda,a}\,=\, (a^{(d-4)/2}/g)^{d(N)\Lambda_r}\,Z_{\Lambda,a}$ as in Eq. (\ref{nZ}). $Z_{\Lambda,a}$ is proved to satisfy thermodynamic and ultraviolet stable stability bounds given in Theorem 1.
%%%%%%%%%%%%%%%%%%%%%%%%%%%%%%&&&&&&&&&&&&&&&&&&&&&&&%%%%%%%%%%%%%%%%%%%%%%%%%%%%%
\section{Upper Stability Bound} \lb{upper}
%%%%%%%%%%%%%%%%%%%%%%%%%%%%%%&&&&&&&&&&&&&&&&&&&&&&&%%%%%%%%%%%%%%%%%%%%%%%%%%%%%
Here, we do not use the enhanced temporal gauge or any other gauge fixing. To prove the upper bound on $Z^n_{\Lambda,a}$ of Eq. (\ref{nZ}), in $\mathcal A=\sum_p\,\mathcal A_p$ appearing in $Z_{\Lambda,a}$ (see Eq. (\ref{Z})), we use the pointwise positivity of $\mathcal A_p$ to discard some horizontal (non-vertical), spatial plaquette actions, corresponding to plaquettes which are orthogonal to the $x^0$-direction. We have
$ e^{-\mathcal A}\leq e^{-\mathcal A_h\,-\,\mathcal A_v}$, where $\mathcal A_v$ is the sum of {\em all} vertical plaquette actions (plaquettes with one side parallel to the time direction) and $\mathcal A_h$ is the sum over the actions of horizontal plaquettes in the plane $x^0=1$, {\em only}. Hence, as given above, recalling that $c\equiv c(a,g^2,d)$, we have \bequ\lb{Zhv}Z_{\Lambda,a}\leq \int \exp[-c(\mathcal A_h\,+\,\mathcal A_v)]\,d\tilde g\,=\, z^{\Lambda_r}\,.\eequ
Even though we have not fixed the gauge, or discarded any gauge bond variables, it is remarkable and surprising that the number of factors of $z$ in Eq. (\ref{Zhv}) , namely $\Lambda_r$ is the same as the number  of retained bonds when the enhanced temporal gauge is fixed (see the discussion above Theorem 1). 

The bound of Eq. (\ref{Zhv}) gives the exact result for $d=2$.  For $d=3$, the right-hand side is $[z(c(a,g,d=3))]^{(2L+1)(L-1)^2}$. 
To understand the contents of Eq. (\ref{Zhv}) in $d=3$ (it is similar for $d=2,4$), we remark that although some plaquette actions have been erased in the exponent of the integrand, we still need to perform the gauge integrals over all the lattice bond variables and that one bond variable may be present in more than one original plaquette.  For this, we observe that the integral in Eq. (\ref{Zhv}) is performed applying the following procedure. First, we treat the integration over the $2L(L-1)$ horizontal bond variables in the hyperplane with $x^0$ taking its maximum value $x^0=L$. Each of these bond variables appears in only one vertical plaquette. Performing the gauge integrations, the integral of each bond variable, in principle, depends on the other variables in the plaquette. But, using the right/left invariance of the Haar measure \cite{Weyl,Simon2}, it is independent of the other bond variables, and we extract a factor $z$. Altogether, we extract a factor $z^{2L(L-1)}$. We are left with the gauge integrals over the vertical bond variables between $x^0=L$ and $x^0=L-1$. But the integrand is independent of these variables and the integral gives 1. Similarly, we treat successive $x^0$ hyperplanes, integrating over their horizontal bond variables. We get a factor of $z^{2L(L-1)^2}$. With this, we are left with the integral over the horizontal bonds of the $x^0$ hyperplane with the smallest $x^0$ value. These integrals can be carried out in various ways. For instance, by integrating over the $(L-1)$ variables in the column between the planes $x^1=1$ and $x^1=2$, starting at $x^1=1$, we extract a factor of $z^{L-1}$. Repeating this procedure over the remaining $(L-2)$ columns we get $z^{(L-1)^2}$. Altogether, we get the factor $z^{2L(L-1)^2+(L-1)^2}=z^{\Lambda_r}$. Following the integration procedure above, it is seen that the horizontal variable integration is exactly over the $\Lambda_r$ retained variables when we fix the enhanced temporal gauge.
%Note that, in the above procedure, we have not used local gauge invariance or gauge fixing. However, the same result does hold using the above integration procedure if we fix the enhanced temporal gauge.

With Eq. (\ref{Zhv}) in mind, the upper bound on $Z_{\Lambda,a}$ is proved by giving an upper bound on $z(c)$ of Eq. (\ref{zz}). For the upper bound of $z(c)$, since the integrand is positive, we cannot restrict the integration domain, as we do below to obtain a lower bound. Instead, we obtain a global quadratic lower bound on the retained actions and a global upper bound on the densities $\rho(\la)$ of the Haar measures. The bounds we obtain are global, i.e. they hold for the whole gauge group $\mathcal G$. For this, we use the lower bound \cite{Simon3},  $2(1-\cos\theta)\geq 4\theta^2/\pi^2$, $|\theta|<\pi$, in the exponent of Eq. (\ref{zz}) and the upper bound $2(1-\cos \theta)\leq \theta^2$,  $\theta\in\mathbb R$, on the density  $\rho(\lambda)$, to get $|e^{i\lambda_j}-e^{i\lambda_k}|^2=2[1-\cos(\la_j-\la_k)]\leq\left(\lambda_j-\lambda_k\right)^2$. Doing this, for $d=3$, with $c(a,g^2,d=3)= (ag)^{-2}$ and setting $y_j=\frac{2\sqrt{2}}{\pi\sqrt{a}g}\lambda_j$, we obtain
\bequ\lb{bbbb}\barr{lll}
z(c(a,g^2,3))&\leq&\mathcal N(N)\displaystyle\int_{[-\pi,\pi]^{N}}\,\exp\left[-\frac{4}{\pi^2ag^2}\sum_{j=1}^{N} \lambda_j^2  \right]%\vspace{2mm}\\&&\times
\,\prod_{1\leq j<k\leq N}\left(\lambda_j-\lambda_k\right)^2\,d^N\lambda
\vspace{2.5mm}\\&\leq&(\sqrt{a} g)^{N^2}\,  [\pi/(2\sqrt{2})]^{N^2}\,I(2\sqrt{2}/\sqrt{a}g)\,,
\earr
\eequ
where we defined
\bequ\lb{I}
I(u)\,\equiv\, {\mathcal N}(N)\,\displaystyle\int_{[-u,u]^N}\!e^{-y^2/2}\!\prod_{1\leq j<k\leq N}\!\left(y_j-y_k\right)^2 d^Ny\,.
\eequ Note that we have extracted, in the above bound on $z(c(a,g^2,3))$, the factor $(\sqrt{a} g)^{N^2}$, which gives a singularity in the free energy in the continuum limit $a\searrow 0$. Note also that the function $I(u)$ is  continuous, monotone increasing, verifies $I(0)=0$ and is bounded from above by $I(\infty)=\mathcal N(N)(2\pi)^{N/2}\prod_{j=1,\ldots,N}\,j! =(2\pi)^{-N/2}\prod_{j=1,\ldots,(N-1)}\,j!$.

Using $I(u)\leq I(\infty)$, Eq. (\ref{bbbb}) and Eq. (\ref{Zhv}), the final upper bound we obtain for the partition function $Z_{\Lambda,a}$ is 
$$
Z_{\Lambda,a}\leq (\sqrt{a} g)^{N^2\Lambda_r}\,\left[[\pi/(2\sqrt{2})]^{N^2}\, \,I(\infty)\right]^{\Lambda_r}\,.
$$

With this, by passing to the normalized partition function $Z^n_{\Lambda,a}$ of Eq. (\ref{nZ}) the upper stability bound $Z^n_{\Lambda,a}\leq e^{c_u\Lambda_r}$ of Theorem 1 is proved with $c_u\geq \ln \{[\pi/(2\sqrt{2})]^{N^2} I(\infty)\}$ and gives an upper bound on the normalized free energy $f^n_{\Lambda,a}$.

We finish this section by observing that, up to a constant factor, the integrand of $I(u)$ is the probability distribution for the eigenvalues of a self-adjoint matrix in the Gaussian Unitary Ensemble (GUE) (see \cite{Metha,Deift}), and arises naturally in the context of our problem.
For $d=2$, the thermodynamic limit of $f^n_{a,\Lambda}$ exists and, by dominated convergence, the continuum limit also exists and is $$f^n=-\ln\sqrt{2}-\dfrac 1{2N}\ln(2\pi)+\dfrac 1{N^2}\sum_{1\leq j\leq(N-1)}\ln (j!)\,.$$
%%%%%%%%%%%%%%%%%%%%%%%%%%%%%%&&&&&&&&&&&&&&&&&&&&&&&%%%%%%%%%%%%%%%%%%%%%%%%%%%%%
\section{Lower Stability Bound} \lb{lower}
%%%%%%%%%%%%%%%%%%%%%%%%%%%%%%&&&&&&&&&&&&&&&&&&&&&&&%%%%%%%%%%%%%%%%%%%%%%%%%%%%%
For the lower bound on $Z_{\Lambda,a}$, we fix the enhanced temporal gauge so that we have $\Lambda_r$ retained bond variables. As the integrand of $Z_{\Lambda,a}$ is positive, we have the luxury to reduce the integration domain to obtain a lower bound. Next, we reduce the integration domain in $Z_{\Lambda,a}$ so that each retained variable is close to the identity. With this reduction, we obtain an upper bound on each plaquette action which is quadratic and local in the gluon fields of its plaquette. The lower bound for $Z_{\Lambda,a}$ factorizes over the retained bond variables. Each factor is a single-bond partition function $\check z$, with a quadratic action and small field restrictions. The single-bond quadratic action is a class function on $\mathcal G$ and the small field restriction is also. Hence, by the Weyl integration formula, the group integral reduces to an integration over angular eigenvalues. A lower bound for the density $\rho(\lambda)$ is used. Last, changing the variables in the retained variables gives the lower stability bound of Theorem 1. With the enhanced temporal gauge, we remark we have the same exponent $\Lambda_r$ of $z$ that occurs in Eq. (\ref{Zhv}).

To discuss the restriction on the domain, we write the $N\times N$ matrix $U_b$, associated with the bond $b$, as $U_b=e^{iX_b}$. Imposing the condition $\mathcal A_b\equiv \|U_b-1\|^2_{H-S}<1$, we have a well-defined self-adjoint $X_b=-i\ln [1+(U_b-1)]=-i\sum_{j\geq 1}\,(-1)^{j+1}\,(U_b-1)^j/j$ where, for $\mathcal G={\rm U}(N)$, $X_b=\sum_{\alpha=1,\ldots,d(N)=N^2} x_\alpha^b\theta_\alpha$, and $\theta_\alpha$ are the corresponding $d(N)=N^2$ Lie algebra generators, verifying $Tr(\theta_\alpha\theta_\beta)= \delta_{\alpha\beta}$, and the $x_\alpha^b$ are the gauge or gluon fields. Using the eigenvalues $\lambda_j$ of $X_b$, $U_b$ is unitarily equivalent to ${\rm diag}(e^{i\lambda_1},\ldots,e^{i\lambda_N})$, and $\mathcal A_b\equiv \|1-U_b\|^2_{H-S}= 2Tr(1-\cos X_b)$. Also,
\bequ\lb{ln}\|X_b\|^2_{H-S}=Tr\{\ln[1+(U_b-1)]^\dagger\,\ln[1+(U_b-1)]\}\,.\eequ
Note that both, $\mathcal A_b$ and $\|X_b\|^2_{H-S}$, are class functions.  From the spectral theorem for a unitary matrix $U_b$, there exists a unitary matrix $V$ such that $V^{-1}U_bV\,=\,\mathrm{diag}(e^{i\la_1},\ldots,e^{i\la_N})$. Thus, $iX_b\,=\,\ln U_b\,=\,iV[\mathrm{diag}(\la_1,\ldots,\la_N)]V^{-1}$.

To discuss the restriction on the domain, we write the $N\times N$ matrix $U_b$, associated with the bond $b$, as $U_b=e^{iX_b}$. Imposing the condition $\mathcal A_b\equiv \|U_b-1\|^2_{H-S}<1$, we have a well-defined self-adjoint $X_b=-i\ln [1+(U_b-1)]=-i\sum_{j\geq 1}\,(-1)^{j+1}\,(U_b-1)^j/j$ where, for $\mathcal G={\mathrm U}(N)$, $X_b=\sum_{\alpha=1,\ldots,d(N)=N^2} x_\alpha^b\theta_\alpha$, and $\theta_\alpha$ are the corresponding $d(N)=N^2$ Lie algebra generators, verifying $Tr(\theta_\alpha\theta_\beta)= \delta_{\alpha\beta}$, and the $x_\alpha^b$ are the gauge or gluon fields. Using the eigenvalues $\lambda_j\in(-\pi,\pi]$ of $X_b$, $U_b$ is unitarily equivalent to ${\mathrm diag}(e^{i\lambda_1},\ldots,e^{i\lambda_N})$, and $\mathcal A_b\equiv \|1-U_b\|^2_{H-S}= 2\mathrm{Tr}\,(1-\cos X_b)$. Also,
\bequ\lb{ln}
\|X_b\|^2_{H-S}=\mathrm{Tr}\,\{\ln[1+(U_b-1)]^\dagger\,\ln[1+(U_b-1)]\}\,.
\eequ
Note that both, $\mathcal A_b$ and $\|X_b\|^2_{H-S}$, are class functions. We introduce the Euclidean norm $|x^b|$ in $\mathbb R^{N^2}$ and the norm $|\la^b|$ in $\mathbb R^N$. From the representation $X_b\,=\,\sum_\alpha x^b_\alpha \theta_\alpha$ and, recalling the orthogonality condition $\mathrm {Tr}\, \theta_\alpha\theta_\beta\,=\,\delta_{\alpha\beta}$, we have
$$
\| X_b\|^2_{H-S}\,=\,\mathrm Tr\left((X^b)^\dagger X^b\right)\,=\,\sum_{\alpha=1}^{N^2}|x_\alpha^b|^2=|x^b|^2\,.
$$
On the  other hand, using the angular eigenvalues $\la^b_1$, ..., $\la^b_N$ of $X_b$, we have
\bequ\lb{id}
\| X_b\|^2_{H-S}\,=\,\sum_{k=1}^{N}(\lambda_k^b)^2 = |\lambda^b|^2\qquad;\qquad|\la^b_k|<\pi\,.
\eequ
Thus, we get the important identity $|x^b|\,=\,|\la^b|$, $|\la^b_k|<\pi$.

Continuing, for the abelian case $\mathcal G={\rm U}(1)$, we have $$\mathcal A_p=2[1-\cos(x^1+x^2-x^3-x^4)]\leq (x^1+x^2-x^3-x^4)^2\leq 4[(x^1)^2+...+(x^4)^2]\,.$$ This global bound may be not true for a nonabelian $\mathcal G$, without a restriction on the fields. To obtain a quadratic upper bound in this case, we take the gluon fields to be small. In the following, we still take $\mathcal G={\rm U}(N)$. For the single plaquette, with subsequent bonds $b_1$, $b_2$, $b_3$ and $b_4$, with action $\mathcal A_p=\|U_1U_2U^\dagger_3U_4^\dagger-1\|_{H-S}^2$, and $U_j=\exp(i\sum_{\alpha=1}^{d(N)}\,x^j_\alpha\theta_\alpha)$, we prove:
\begin{lemma}
Let $U_p=U_1U_2U_3^\dagger U^\dagger_4$, $U_j=e^{\mathcal L_j}$, $1\leq j\leq 4$ and $\mathcal L_j=i\sum_{\alpha=1,\ldots,d(N)} x^j_\alpha\theta_\alpha$. Then,\\
\noindent a) if $\|\mathcal L_j\|_{H-S}<N^{-1/2}$, we have
$$\barr{lllll}
\!\mathcal A_p&\leq& 4\left( 1+\!N^2\,\sum_{j=1}^4\|\mathcal L_j\|_{H-S}+\!N^4\, \sum_{j=1}^4\|\mathcal L_j\|_{H-S}^2 \right)\,\sum_{j=1}^4\|\mathcal L_j\|_{H-S}^2%\\&&
&\leq& C^2\,\!\sum_{k=1}^4|x^k|^2\,,\earr
$$
where $C=2(1+2N^{3/2})$ and $|x^j|=\|\mathcal L_j\|_{H-S}$.\\
\noindent b) The total action $\mathcal A\,=\,\sum_p\,\mathcal A_p$ obeys the quadratic upper bound
$$
\mathcal A\,\leq\,2(d-1)\, C^2\,\sum_b\,|x^b|^2\qquad,\qquad |x^b|<N^{-1/2}\,,
$$
where the sum is over the $\Lambda_r$ retained bonds only.
\end{lemma}

The proof of Lemma 1 is given in the end of the section.
Now, we obtain a lower bound on $Z_{\Lambda,a}$ in terms of the modified single-bond partition function $\check z$. We first obtain a condition on $\|X_b\|_{H-S}=|x^b|$ so that with $U_b=e^{iX_b}$, we have $\| U_b-1\|_{H-S}<1$. This guarantees that there is a $1-1$ correspondence between $x_b$ and its range under the exponential map. By Taylor expanding $U_b(\alpha)\equiv e^{\alpha\mathcal L_b}$, $(U_b-1)=\int_0^1\mathcal L_bU_b(\alpha)d\alpha$ and $\|U_b-1\|_{H-S}\leq\|\mathcal L_b\|_{H-S}\,\sup_{\alpha\in[0,1]}\|U_b(\alpha)\|_{H-S}\leq N^{1/2}\|\mathcal L_b\|_{H-S}$. Here, for a unitary $U$, we have used $\|U\|_{H-S}=N^{1/2}$.

The logarithms in Eq. (\ref{ln}) are defined if, for the $\Lambda_r$ retained bonds, we impose $|x^b|<(1/\sqrt{N})$, so that $\|U_b-1\|_{H-S}<1$. Lemma 1 applies and, in $\mathcal A\!=\!\sum_p\!\mathcal A_p$, we replace $\mathcal A_p$ by the quadratic bound. % in $\mathcal A=\sum_p\,\mathcal A_p$.
Since the integrand is now a class function of each retained %of the $\Lambda_r$
bond variable, we replace the integration variables by the angular eigenvalues with $|\lambda^b_k|<(1/N)\equiv\gamma$. With this condition, $|\la^b|=(\sum_{k=1,\ldots,N}\,|\la^b_k|^2)^{1/2}<1/\sqrt{N}$ and, by the identity $|x^b|=|\la^b|$, we have $|x^b|< 1/\sqrt{N}$. Each bond $b$ appears at most in $[2(d-1)]$ terms of $\mathcal A$. Hence, paying with a factor $[2(d-1)]$, $\sum_p$ is replaced by the sum over retained bonds. With this, the bound on $Z_{\Lambda,a}$ factorizes over the retained bonds to give (compare with Eq. (\ref{zz})),
$$\!
Z_{\Lambda,a}\geq \check z^{\Lambda_r}=\left[\mathcal N(N)\,\!\int_{|\lambda_k|<\gamma}e^{-2c(d-1)C^2\sum_{k=1}^N \lambda^2_k}\rho(\lambda)d^N\lambda\right]^{\Lambda_r}\,\!\!,
$$
Recall that, $c\equiv c(a,g^2,d)$. For the lower bound on
$Z_{\Lambda,a}$, we use a lower bound on the eigenvalue density $\rho(\lambda)$. Namely, we use $|e^{i\lambda_j}-e^{i\lambda_k}|^2=2[1-\cos(\la_j-\la_k)]\geq 4\left(\lambda_j-\lambda_k\right)^2/ \pi^2$, $|\lambda_\ell|\leq \pi/2$. After the change of variables $y_k=[4c(d-1)]^{1/2}C \lambda_k$, we get
$$
Z_{\Lambda,a}\geq  (g^2/a^{d-4})^{d(N)\Lambda_r/2}\left[ \Theta \,I(2(d-1)^{1/2}C\gamma a^{(d-4)/2}/g^2)\right]^{\Lambda_r}\,,
$$
with $\Theta\equiv \mathcal N(N)(4/\pi^2)^{N(N-1)/2}[(d-1)^{1/2}2C\gamma]^{-d(N)}\!$ which displays the factor $(g^2/a^{d-4})^{d(N)/2}$. With the free energy ($\propto \ln Z_{\Lambda,a}$) in mind, we emphasize this is the same singular factor which occurs above Eq. (\ref{I}) for the upper bound of $z$ and, by the monotonicity of $I(u)$, the integral is bounded below for $a=1$ and $g^2=g^2_0<\infty$. Letting $I_\ell$ denote the value of $I(\cdot)$ for $a=1$ and $g^2=g_0^2$, we have the lower bound $$Z_{\Lambda,a}\geq\left(g^2/a^{d-4}\right)^{d(N)\Lambda_r/2} \;\left( \Theta \,I_\ell\right)^{\Lambda_r}\, \geq\,\left(g^2/a^{d-4}\right)^{d(N)\Lambda_r/2}\,e^{c_\ell d(N) \Lambda_r}\,,$$
for any $c_\ell \,\leq\, [d(N)]^{-1}\ln (\Theta I_\ell)$, which gives the lower stability bound on the normalized partition function $Z^n_{\Lambda,a}$ of Theorem 1.

We now write $|\cdot|$ for $\|\cdot\|_{H-S}$ and prove Lemma 1.

\noindent{\bf Proof of Lemma 1:} We first prove item a) of Lemma 1. For $\delta\in[0,1]$, we let $U_p(\delta)=U_1(\delta)U_2(\delta)U^\dagger_3(\delta)U^\dagger_4(\delta)$, where $U_j(\delta)=e^{\delta\mathcal L_j}$. Clearly, $U_j=U_j(\delta=1)$. With $\mathcal L=\sum_{j=1,\ldots,4} \mathcal L_j$ and applying a Taylor expansion, we have
$$
U_p\!=\!1\!+\!\mathcal L\!+\!\int_0^1\!\int_0^{\delta_2}[d^2U_p(\delta_1)/d\delta_1^2]\,d\delta_1 d\delta_2\equiv 1\!+\!\mathcal L\!+\!\mathcal R\,,
%\earr
$$
Since $(dU_j/d\delta)\!=\!U_j\mathcal L_j\!=\!\mathcal L_jU_j$ and suppressing the $\delta$'s, we have $[d^2U_p(\delta)/d\delta^2]=\mathcal L_1\mathcal L_1U_p+\mathcal L_1U_1\mathcal L_2U_2U^\dagger_3U^\dagger_4+\ldots+U_p\mathcal L_4\mathcal L_4$.
Thus, since $|U_j|=N^{1/2}$ and $2ab\leq a^2+b^2$, $a,b\in\mathbb R$, we obtain $|\mathcal R|\leq (N^2/2) \sum_{j,k}|\mathcal L_j|\,|\mathcal L_k|\leq 2N^2\sum_j|\mathcal L_j|^2\equiv 2N^2\mathcal Q$. With the small field condition,
$$\barr{lllll}\!\mathcal A_p&=&|U_p-1|^2=|\mathcal L+\mathcal R|^2\leq\, (|\mathcal L|+|\mathcal R|)^2&=&|\mathcal L|^2+2|\mathcal L||\mathcal R|+|\mathcal R|^2\,.\earr$$
But $|\mathcal L|^2\leq(\sum_j\,|\mathcal L_j|)^2
\leq 4 \sum_j\,|\mathcal L_j|^2$, by the Cauchy-Schwarz inequality. So,

$$\barr{lllll}\!\mathcal A_p&\leq&
4\left(1+N^2\sum_j|\mathcal L_j|  +N^4\mathcal Q\right)\mathcal Q\,\leq\, C^2\,\sum_{k=1}^4|x^k|^2\,.\earr$$

With this upper bound on $\mathcal A_p$, the upper bound on the total action $\mathcal A$ arises from summing over the $\Lambda_r$ retained plaquettes, noting that a bond may be present in at most $2(d-1)$ plaquettes.\qed

The above discussion implies the lower stability bound of Theorem 1 for the case of the gauge group ${\mathrm U}(N)$, and leads to a lower bound on the normalized free-energy $f^n_{\Lambda,a}$. The proof of Theorem 1 is finished in the next section where we extend our results to $\mathcal G={\mathrm SU}(N)$.
%%%%%%%%%%%%%%%%%%%%%%%%%%%%%%&&&&&&&&&&&&&&&&&&&&&&&%%%%%%%%%%%%%%%%%%%%%%%%%%%%%
\section{Extension of Results to ${\rm SU}(N)$} \lb{extension}
%%%%%%%%%%%%%%%%%%%%%%%%%%%%%%&&&&&&&&&&&&&&&&&&&&&&&%%%%%%%%%%%%%%%%%%%%%%%%%%%%%
Our analysis and results extend from the gauge group $\mathcal G={\rm U}(N)$ to $\mathcal G={\rm SU}(N)$. To do this, besides noticing there are now $d(N)=(N^2-1)$ self-adjoint and traceless Lie algebra generators, we make the following changes:\vspace{1mm}\\
\noindent {\bf a)} Using a Dirac delta, in the Weyl integration formula, insert $[2\pi\delta(\xi)]$  in the integrand, with $\xi\equiv\lambda_1+\ldots+\lambda_N$;\vspace{1mm}\\
\noindent {\bf b)} For the lower bound of $\sum_{k=1,...,N}[2(1-\cos\lambda_k)]$, $\xi=0$, replace $N$ by $(N-1)$ and use $\sum_{k=1,...,N-1}[2(1-\cos\lambda_k)]\geq \sum_{k=1,\ldots,(N-1)}[2\lambda_k^2/\pi^2]$; $|\lambda_k|<\pi$. For the upper bound of $\sum_{k=1,...,N-1}[2(1-\cos\lambda_k)]+2[1-\cos(\xi-\lambda_N)]$, use $[N(\lambda_1^2+\ldots+\lambda_{N-1}^2)]$;\vspace{1mm}\\
\noindent {\bf c)} For the upper bound on the density, we have $\rho(\lambda)\leq\check\varrho(\lambda)$, where, for $j,k=1,\ldots,(N-1)$, $\check\varrho(\lambda)\!\equiv\! \prod_{j<k}(\lambda_j-\lambda_k)^2\prod_{j}(\lambda_j+\xi-\lambda_N)^2$. For the lower bound, restrict the set $\{\lambda_k\}$ so that we can use $(1-\cos u)\geq2u^2/\pi$, $|u|<\pi$. Taking $|\lambda_k|\leq\pi/N$, we have $\rho(\lambda)\geq (2/\pi^2)^{N(N-1)/2}\check\varrho(\lambda)$.

With another convenient $c_\ell$ for the ${\mathrm SU}(N)$ case, and taking the smallest between the two $c_\ell$ for $\mathcal G={\mathrm U}(N),{\mathrm SU}(N)$,  we conclude the proof of Theorem 1.\qed
%%%%%%%%%%%%%%%%%%%%%%%%%%%%%%&&&&&&&&&&&&&&&&&&&&&&&%%%%%%%%%%%%%%%%%%%%%%%%%%%%%
\section{Concluding Remarks} \lb{conclusion}
%%%%%%%%%%%%%%%%%%%%%%%%%%%%%%&&&&&&&&&&&&&&&&&&&&&&&%%%%%%%%%%%%%%%%%%%%%%%%%%%%%
We consider a pure Yang-Mills model in dimension $d = 2,3,4$, with partition function $Z_{\Lambda,a}$ defined, with the Wilson plaquette action and with a compact and connected Lie group $\mathcal G$, on a hypercubic finite lattice $\Lambda\subset a\mathbb Z^d$, $a\in(0,1]$, with $L\in\mathbb N$ sites on a side. %This lattice has $\Lambda_s=L^d$ sites and $\Lambda_b=dL^{d-1}(L-1)$ bonds.
We take $\mathcal G\,=\,{\rm U}(N),\,{\rm SU}(N)$ and  $d(N)$ is the dimension of the corresponding Lie algebras. The lattice provides an ultraviolet, short-distance regularization.

In a series of papers (see \cite{Bal,Bal2} and Refs. therein), ultraviolet stable stability bounds were proved for the $d=3,4$ pure-gauge cases, using intricate rigorous renormalization group methods and field decompositions.

In our treatment, the gauge group % $\Lambda_b$
bond variables are parametrized using $d(N)$ gluon fields. By local gauge invariance, sometimes it is convenient to implement a gauge fixing procedure which allows us to gauge away some bond variables by setting them equal to the identity group element in the Wilson plaquette action. The corresponding gauge group integral gives one. This procedure can be carried out without changing the value of the finite lattice model partition function $Z_{\Lambda,a}$. We use an enhanced temporal (axial) gauge where the temporal bond variables are set to one; some additional bond variables on the lattice boundary are also set to one. The set of gauged away bonds (variables) are associated with a maximal tree with bonds of $\Lambda$. A closed loop is formed if any other bond is added and gives a different partition function. The number of retained bond variables is denoted by $\Lambda_r$, and is of order $(d-1)L^d$, for large $L$.

By extracting a factor $(a^{d-4}/g^2)^{-d(N)\Lambda_r/2}$ from $Z_{\Lambda,a}$, so that
$Z_{\Lambda,a}= (a^{d-4}/g^2)^{-d(N)\Lambda_r/2}\,Z^n_{\Lambda,a}$, we show that the normalized partition function $Z^n_{\Lambda,a}$ obeys the thermodynamic and ultraviolet stable, stability bounds
$$\exp\left[c_\ell d(N)\Lambda_r\right]\,\leq\, Z^n_{\Lambda,a}\,\leq\, \exp\left[c_ud(N)\Lambda_r\right]\,,$$ with finite $c_\ell,\,c_u\in\mathbb R$ independent of $L$, $a$ and the gauge coupling $g^2\in(0,g_0^2]$, $0<g_0<\infty$. In other words, we extracted the exact singular behavior of the finite lattice free-energy $f_{\Lambda,a}= \ln Z_{\Lambda,a}/[d(N)\Lambda_r]$ and the normalized free-energy $f^n_{\Lambda,a}= \ln Z^n_{\Lambda,a}/[d(N)\Lambda_r]$ has both a thermodynamic limit ($\Lambda\nearrow\infty$), at least in the subsequential sense, and subsequently a continuum limit ($a\searrow 0$).

Both the upper and the lower bounds on $Z^n_{\Lambda,a}$ factorize and reduce to the $d(N)\Lambda_r$ power of a single plaquette, single bond partition function $z$ (for the lower bound, with a small gluon field restriction). The integrand of $z$ is a class function on $\mathcal G$, and we apply the Weyl integration formula to perform the gauge field integrals. The bounds reduce to bounds on the angular eigenvalue distribution of unitary matrices and arise from quadratic bounds on the action, with a restriction on the size of the fields for the lower bound. This appears to indicate that the apparently high nonlinearities of gauge models are actually not that bad. Our bounds give the exact result for $d=2$.

Extending the stability bounds to partition functions with a uniform source coupled with a sequence of well-known techniques (e.g., multiple reflection \cite {GJ}), may lead to the thermodynamic and continuum limits of correlations.  More analysis is needed for correlation decays.

Our treatment extends to other connected and compact groups $\mathcal G$, and when matter fields are present. Indeed, neglecting the pure-gauge action and using a priori locally scaled matter fields (not canonical scaling!) the coupling of matter and gauge fields was treated in \cite{M,MP,MP2}. Stability bounds were proven for a  Bose-gauge model; only upper bounds in the Fermi case. The bounds do not depend on $a$ and a normalized free-energy exists in the thermodynamic and continuum limits. We expect to combine our pure-gauge and matter-gauge results to show the existence of QCD. Scalar QCD models, with scalar fields replacing the quarks, are analyzed in \cite{bQCD}.

Stability bounds give the existence of the model but do not give information on the energy-momentum spectrum, local clustering properties and particles \cite{GJ}. For lattice QCD, with fixed $a$ and in the strong coupling regime (with $g^{-2}>0$ much smaller than a small hopping parameter), we have results validating the Gell'Mann-Ne'eman eightfold way, the exponential decay of the Yukawa interaction and the existence of some two-hadron bound states (see e.g. \cite{WPT-Baryons,WPT-Mesons} and references therein). It would be nice and desirable to rigorously derive general properties of nuclear physics from first principles, i.e. from fundamental quarks and gluons, and QCD dynamics.
%%%%%%%%%%%%%%%%%%%%%%%%%%%%%%%%%%%%%%%%%%%%%%%%%%%%%%%%%%%%%%%%%%%%%%%%%%%%
%%%%%%%%%%%%%%%%%%%%%%%%%%%%%%%%%%%%%%%%%%%%%%%%%%%%%%%%%%%%%%%%%%%%%%%%%%
\begin{acknowledgements}
We would like to acknowledge the partial support of Funda\c c\~ao de Amparo \`a Pesquisa no Estado de S\~ao Paulo (FAPESP) and Conselho Nacional de Desenvolvimento Cient\'\i fico e Tecnol\'ogico (CNPq).
\end{acknowledgements}
%%%%%%%%%%%%%%%%%%%%%%%%%%%%%%%%%%%%%%%%%%%%%%%%%%%%%%%%%%%%%%%%%%%%%%%%%%%%
%%%%%%%%%%%%%%%%%%%%%%%%%%%%%%%%%%%%%%%%%%%%%%%%%%%%%%%%%%%%%%%%%%%%%%%%%%

\end{document}